The highly peculiar emission lines detected in spectra of extragalactic objects may be generated by ultra-rapid quasi-periodic oscillations

Ermanno F. Borra

Département de Physique, Université Laval, Québec, Canada

(borra@phy.ulaval.ca)

Abstract

Extremely peculiar emission lines have been found in the spectra of some active galactic nuclei and quasars. Their origin is totally unknown. We investigate the hypothesis that they are generated from ultra-rapid quasi-periodic oscillations that may occur in jets or black holes, as predicted in a published theoretical paper. We conclude that, although not totally certain, this hypothesis is just as valid as the other highly peculiar hypotheses that have been previously made (e.g. blueshifts due to bulk motions close to the speed of light). We consider ways to further validate our hypothesis.

1. Introduction

The time domain is by far the least explored of all of the Astronomical physical domains. The difficulty with the short time domain is that it requires specialized and expensive instrumentation. It also requires extremely large quantities of data that are difficult to store and analyze. Furthermore, extremely fast time variations (e.g. $10^{-11}$ seconds) cannot even be detected with currently available instrumentation. Borra (2010) proposed a very simple technique to find extremely rapid periodic time variations that have periods shorter than $10^{-11}$ seconds. Pulses separated by times shorter than $10^{-11}$ seconds generate periodic modulations in optical spectra that can be easily detected by carrying out Fourier transforms of the spectra. Borra (2013) discusses the results of a Fourier transform analysis of the spectra of 0.9 million galaxies and 0.5 million stars and quasars in the Sloan Digital Sky Survey (SDSS). The Fourier transform analysis detected ultra-rapid periodic modulations in the spectra of only 223 galaxies. There is no doubt that the signals are real, and not caused by instrumental effects, because plots of the periods as a function of redshift show that the periods increase with redshifts following two tight linear relations. These linear increases with redshift imply two base time periods of $1.02 \times 10^{-13}$ seconds and $1.09 \times 10^{-13}$ seconds in the reference frames of the galaxies. Furthermore, none of the stars had signals at those time periods, while most of the stars had brighter apparent magnitudes than the galaxies and therefore gave spectra that had higher signal to noise ratios. Note that the fact that the galaxies are extended sources, while the stars are not, does not have any impact because the SDSS spectra used by Borra (2013) observe the cores of the galaxies with optical fibers that have a 3 arcseconds diameter.

Borra (2017) extended the theoretical analysis of Borra (2010) to the case where the pulses are quasi-periodic. Quasi-periodic pulses are separated by the same average time when the average is carried out over long times; however the time separation between individual pulses varies at random. Computer simulations in Borra (2017) of the effect of quasi-periodic pulses show that they generate very narrow spectral lines if the average period is constant over the observation time. The lines are broadened if the average period is not constant over the observation time.

Extremely peculiar spectral emission lines have been found in the optical spectra (Mathee et al. 2017) and X-ray spectra (Yaqoob et al. 1998) of distant active galactic nuclei (AGNs) and quasars. The origin of these lines is totally unknown, but Yaqoob et al. (1998) hypothesize that they come from extremely blueshifted known spectral lines. They assume that the blueshift is due to extremely large bulk velocities (e.g. 0.75 the speed of light) or gravitational shifts from supermassive black holes. Several surveys were carried out to find Lyman alpha emitters. For example Mathee et al. (2017) observed several luminous Lyman alpha emitters. They found other spectral lines that confirm the Lyman alpha assumption in all of them except two of them,

In this work we will consider the hypothesis that the discovered peculiar emission lines are generated from ultra-rapid quasi-periodic oscillations.

2. Summary of the theory of generation of emission lines from quasi-periodic oscillations

This section contains a brief summary of the theory discussed in Borra (2017) that shows that ultra-rapid quasi-periodic pulses generate spectral features that look like emission lines.

Borra (2017) starts from the theory in Borra (2010) that models the time variations of an electric field $E(t)$, emitted by a source that sends $N$ periodic electric field pulses $V(t)$ that have a period $\tau$, by the convolution of $V(t)$ with a comb function $\sum_m \delta(t - t_m)$ made of the sum of delta functions $\delta(t - t_m)$ separated by $t_m = (m-1)\tau$, where $m$ an integer number. The electric field $E(t)$ is then given by the convolution of $V(t)$ and $\sum_m \delta(t - t_m)$

$$E(t) = V(t) \otimes \sum_m \delta(t - t_m) \quad . \tag{1}$$

The Fourier transform of the intensity, given by the square of the electric field in equation 1, gives a spectrum as a function of frequency. Borra (2010) then shows that a quadratic detector, like a CCD, in a spectrograph detects a spectroscopic modulation given by

$$S(\omega) = S_1(\omega)\left[\sin(\omega N\tau/2)/(\sin(\omega\tau/2))\right]^2 \quad , \tag{2}$$

where $\omega = 2\pi\nu$ and $\nu$ is the frequency in the spectrum. This theoretical analysis is validated by the laboratory experiments of Chin et al. (1992). Borra (2010) then models the electric field $E(t)$ of a very large number of periodic pulses separated by the constant period $\tau$ by the convolution of the comb function with a Gaussian that models the shapes of the individual pulses. The Fourier transform of this signal gives a frequency spectrum made of very narrow spectral lines, separated by a constant frequency separation $\Delta\nu =$

*1/τ*, having strengths that decrease with increasing frequency. Borra (2017) starts from this analytical model to carry out numerical simulations of a Shah function where the separation between the pulses varies with random deviations from the period *τ*. Figure 4 in Borra (2017) shows the spectrum generated by this model and figure 3 in Borra (2017) shows the quasi-periodic electric field that generates this spectrum. The spectrum in figure 4 in Borra (2017) is dominated by a very strong and narrow emission line, while the other lines are weakened below detection. The numerical modeling in Borra (2017) therefore demonstrates that very rapid quasi-periodic time variations generate a spectral feature that would be identified as an emission line in an astronomical spectrum. Because this emission line is generated at a location set by the average period *τ*, it would probably be located at a peculiar spectral location and therefore be identified as a very peculiar spectral line.

3. Generation from quasi-periodic oscillations of peculiar emission lines found in surveys

Several objects have peculiar emission lines that have a totally unknown origin. For example Matthee et al. (2017) discuss the results of their spectroscopic follow-up of a large number of candidate luminous Lyman alpha emitters (LAEs) in the redshift range $5.7 < z < 6.6$. They discuss the spectra of 2 very peculiar LAEs (SR6 and VR7). They only find a strong emission line in their spectra, which they assume to be Lyman alpha at redshifts of $z = 5.676$ for SR6 and $z = 6.532$ for VR7. However, while they searched for other spectral emission or absorption lines at the redshifts implied by the Lyman alpha assumption, they did not find any other line that validates the assumption. They clearly state that SR6 and VR7 are their most luminous objects, VR7 being the most luminous one, but do not have other spectral lines despite their luminosities; consequently the Lyman alpha assumption is only a hypothesis. Sobral et al. (2015) also find a very peculiar LAE (MASOSA) that only has a very strong emission line but no other spectral lines and an undetected continuum. They compare them to Himiko (Ouchi et al. 2009, 2013) which also has a similar spectrum with a single strong emission line but no other spectral features. Therefore there are several cases of astronomical objects that only have a single strong emission line but no other spectral line to confirm the Lyman alpha assumption.

Yaqoob et al. (1998) found a peculiar narrow emission-line spectral feature in the X-Ray spectrum of the quasar PKS 0637-752 which is at the redshift z = 0.654. This narrow emission-line is positioned at a spectroscopic energy location of 1.6 keV in the quasar frame. Note that this 1.6 keV energy is not the energy contained in the line. The energy in keV units is commonly used to identify the spectral location of the line in an X-Ray spectrum, while the wavelength is used in optical spectra. A spectral line had never been found at that X-Ray spectral location before. Yaqoob et al. (1998) reject the hypothesis that it is caused by instrumental effects and explore other possible causes of this spectral line. Shortly after, Yaqoob et al. (1999) found an emission line in the radio-loud quasar PKS 2149 –306 at an X-Ray spectral energy location of 17 keV in the quasar frame which is at a redshift z = 2.345. They conclude that the most likely explanation of the spectral line comes from the hypothesis that it is a Fe-K emission which is blueshifted by a bulk velocity of 0.75c, where c is the speed of light. Since then, several detections of peculiar emission-lines in X-ray spectra have been found (e.g. Turner et al 2010, Bottacini et al. 2014). They are commonly assumed to be caused by known spectral lines (e.g. Fe-K emission line) that are extremely blueshifted or redshifted by extremely high gravitational or bulk velocity shifts.

We will now investigate the hypothesis that these peculiar emission lines are caused by an extremely rapid quasi-periodic oscillation (QPO). The conclusion in Borra (2017) discusses the criteria needed to check the validity of a QPO spectral line. Although the discussion in Borra (2017) is centered on the spectra of Lyman α emitters, the same criteria apply to any spectral domain. One should simply look for other spectral features (e.g spectral lines or continuum breaks) at the same redshift or blueshift implied by the redshift or blueshift of the hypothetical known emission line (e.g. Fe-K emission). The presence of other spectral features would obviously invalidate the QPO hypothesis. The spectra of the objects in which a peculiar emission line has been found (Turner et al 2010, Bottacini et al. 2014, Matthee et al.2017, Sobral et al. 2015, Ouchi et al. 2009, 2013) do not have other spectral features at the redshifts implied by the hypotheses that are commonly made (e.g. a blue-shifted Fe-K line). Consequently, this leaves the QPO hypothesis open.

4. Discussion

Quasi-periodic oscillations may occur in exotic objects like black holes or jets in AGNs and quasars. Borra (2013) carries out an analysis that concludes that the extremely rapid periodic oscillations detected, by Fourier transforms of spectra, in the cores of galaxies are generated by jets in the core of the galaxies.

Quasi-periodic oscillations have already been found in a few black hole binaries that have masses of a few solar masses (Remillard & McClintock 2006). They have average periods as short as 3 milliseconds. The physics responsible for these quasi-periodic oscillations is unknown and many theoretical explanations have been proposed (see the introduction in Varniere & Vincent 2016 for a summary). The fact that rapid quasi-periodic oscillations have already been found in these exotic astronomical objects, although it obviously does not entirely validate it, strengthens the hypothesis that the emission lines found in several AGNS and quasars could be generated by extremely rapid oscillations. Supermassive black hole binaries, that have masses of the order of $10^8$ solar masses, have been found (Rodriguez et al. 2006, Boroson & Lauer 2009, Deane et al. 2014) at the center of active galactic nuclei. It is therefore conceivable that these supermassive binary black holes may generate the kind of ultra-rapid quasi-periodic pulses that we assume generate the emission lines. Because the physics responsible is unknown, it is conceivable that supermassive black holes may generate quasi-periodic oscillations that have average periods far shorter than those found in binary black holes that have masses of the order of the mass of the sun (Remillard & McClintock 2006).

The discussion in the previous paragraph is centered on black holes but, as mentioned at the beginning of this section, quasi-periodic oscillations may also occur in jets. Anything is possible in the relativistic jets that are in quasars and the cores of AGNs and the discovery of ultra-rapid (1.02 x $10^{-13}$ seconds and 1.09 x $10^{-13}$ seconds) periodic modulations in the cores of 223 galaxies by Borra (2013) shows that relativistic jets can generate extremely rapid oscillations.

A QPO emission line is located, in a spectrum, at a frequency $v = 1/\tau$ (Equation 2), where $\tau$ is the average period of the pulsations, so that a QPO generated spectral line in the X-ray spectral region at $10^{18}$ Hz would imply a value of the average period $\tau = 10^{-18}$ seconds (see the discussion in section 2 and Equation 2). This value of the average period $\tau$ seems, at first sight, too short to make sense since it would require extremely high energy density values. However, the discovery of spectroscopic modulations coming from periodic pulses that have periods of $10^{-13}$ seconds in the cores of a small number of galaxies (Borra 2013) shows that pulses having extremely short periods can be generated. The quasi-periodic oscillations coming from black hole binaries, having a few solar masses, that have already been found, also require extremely high energy density values. Furthermore, extremely short unresolved pulses (<0.5 ns) have been detected in the Crab pulsar (Hankins & Eilek 2007). They have a similar problem since this implies a brightness temperature higher than $2 \times 10^{41}$ K.

One may, of course, wonder about what physical phenomena can possibly generate pulses separated by such extremely short times. The answer that we can give to this question is that anything is possible in jets or the extremely massive black holes that have been found at the center of many galaxies and that the discovery of quasi-periodic oscillations with time scales shorter than 2 milliseconds in binary black holes (Remillard & McClintock 2006) strengthens the QPO hypothesis. Although there is no evidence that the average periods of the quasi-periodic oscillations having average periods shorter than 2 milliseconds, presently found in a small number of black holes that have masses of a few solar masses, decrease with increasing mass, it is conceivable that the time scales of the quasi-periodic oscillations decrease considerably in the supermassive black holes, that have masses as large as thousands of billions of solar masses present at the center of galaxies. Emission lines may therefore be generated by ultra-rapid quasi-periodic oscillations in the kind of supermassive binary black holes that have been discovered in Galaxies. This discussion about the masses of black holes is irrelevant, if the quasi-periodic oscillations occur in jets, like the extremely rapid oscillations discussed in Borra (2013).

5. Conclusion

At this stage, the hypothesis that the spectral lines in the spectra of AGNs and quasars are generated by ultra-rapid quasi-periodic oscillations has no greater validity than the other hypotheses discussed in Yaqoob et al. (1999) and Bottacini et al, (2014). However note that, while the fact that the spectra of the objects in which a peculiar emission line has been found do not have other spectral features at the redshifts implied by the hypothesis made by Yaqoob et al. (1999) and Bottacini et al, (2014) weakens their hypothesis, it strengthens the QPO hypothesis, The QPO hypothesis must be validated with further work. In principle, it should be validated with observations of the intensity as a function of time; however, time scales of the order of $10^{-18}$ seconds are too short to be detected with existing instrumentation. Presently, the only way to strengthen the QPO hypothesis is to obtain more spectra with very high signal to noise ratios. As discussed in section 3, one should simply look for other spectral features at the redshift of the emission line since their detection would invalidate the QPO hypothesis. However, while the non-detection of other spectral features would strengthen the hypothesis, it may not totally validate it, since these spectral features may be present but not be detected simply because they are too weak. The strength of these spectral features could be computer modelled to verify whether they are too weak to be detected is valid or not.

At first sight, the QPO hypothesis may seem very strange but, on the other hand, the other hypotheses presently considered (e.g. blueshifts at nearly the speed of light) are also highly peculiar. The fact that quasi-periodic oscillations having average periods as short as 3 milliseconds have already been found in black hole binaries (Remillard & McClintock 2006) gives some credibility to the QPO hypothesis. Finally let us remember that supermassive black hole binaries at the center of AGNs have also been found (Rodriguez et al. 2006, Boroson & Lauer 2009, Deane et al. 2014) and that it is possible that supermassive binary black holes may generate far more rapid oscillations. As mentioned several times in this paper, it is also possible that quasi-periodic oscillations may occur in jets and not black holes. The final conclusion is therefore that, although not certain, the QPO hypothesis is just as valid as the other hypotheses. More observations followed by computer modelling are needed to fully validate it.